\documentclass[a4paper,prx,twocolumn,groupedaddress,showpacs]{revtex4}

\usepackage[utf8]{inputenc}
\usepackage{amssymb}
\usepackage{amsmath}
\usepackage{color}
\usepackage{graphicx}
\usepackage{bbold}
\usepackage{nicefrac}
\usepackage{mathcomp}
\usepackage{subfigure}
\usepackage[colorlinks=True, linkcolor=blue, citecolor=blue, urlcolor=blue]{hyperref}
\usepackage{color}
\usepackage{soul}

\newcommand{\kfeas}{KFe$_2$As$_2$}

\begin{document}

\author{Steffen Backes}
\affiliation{Institut f\"ur Theoretische Physik, Goethe-Universit\"at Frankfurt, Max-von-Laue-Str. 1, 60438 Frankfurt am Main, Germany}
\author{Daniel Guterding}
\affiliation{Institut f\"ur Theoretische Physik, Goethe-Universit\"at Frankfurt, Max-von-Laue-Str. 1, 60438 Frankfurt am Main, Germany}
\author{Harald O. Jeschke}
\affiliation{Institut f\"ur Theoretische Physik, Goethe-Universit\"at Frankfurt, Max-von-Laue-Str. 1, 60438 Frankfurt am Main, Germany}
\author{Roser Valent\'i}
\affiliation{Institut f\"ur Theoretische Physik, Goethe-Universit\"at Frankfurt, Max-von-Laue-Str. 1, 60438 Frankfurt am Main, Germany}

\date{\today}


\title{Electronic structure and de Haas-van Alphen frequencies in {\kfeas} within LDA+DMFT}

\begin{abstract}
  Recent density functional theory (DFT) calculations for {\kfeas}
  have shown to be insufficient to satisfactorily describe
  angle-resolved photoemission (ARPES) measurements as well as
  observed de Haas van Alphen (dHvA) frequencies. In the present work,
  we extend DFT calculations based on the full-potential linear
  augmented plane-wave method by dynamical mean field theory
  (DFT+DMFT) to include correlation effects beyond the local density
  approximation.  Our results indicate that {\kfeas} is a moderately
  correlated metal with a mass renormalization factor of the Fe $3d$
  orbitals between 1.6 and 2.7. Also, the obtained shape and size of
  the Fermi surface are in good agreement with ARPES measurements and
  we observe some topological changes with respect to DFT calculations
  like the opening of an inner hole cylinder at the $Z$ point.  As a
  result, our calculated dHvA frequencies differ greatly from existing
  DFT results and qualitatively agree with experimental data. On this
  basis, we argue that correlation effects are important to understand
  the -presently under debate- nature of superconducting state in
  {\kfeas}.
\end{abstract}

\maketitle

\section{INTRODUCTION}
\label{sec:introduction}
{\kfeas} is the hole-doped end member of the well-studied
Ba$_{1-x}$K$_x$Fe$_2$As$_2$ family of iron based
superconductors~\cite{rotter2008} and it features superconductivity at
$T_c = 3.4 \, \mathrm{K}$ without the need for application of external
pressure~\cite{kihou2010}. This material is presently under debate
since the origin of the superconducting phase and the pairing symmetry
are still unclear~\cite{wang2011,hirschfeld2011,chubukov2012}, both on
the experimental and theoretical sides: Recent laser-based
angle-resolved photoemission (laser ARPES) measurements found the
superconducting order parameter to be of $s$-wave character with octet
line-nodes~\cite{okazaki2012}, while a theoretical
study~\cite{thomale2011} based of functional renormalization group
considerations predicted a $d$-wave symmetry in agreement with
measurements of thermal conductivity~\cite{reid2012}. However, other
theoretical studies~\cite{maiti2011, suzuki2011} based on spin pairing
theory within the random phase approximation  found  that
$s$- and $d$-wave pairing channels are strong competitors in
electron-doped Fe-based systems and, therefore, both are possible in
{\kfeas}. Moreover, transport measurements under
pressure~\cite{tafti2013} suggested the presence of a possible phase
transition from $d$-wave to $s$-wave around $1.75$~GPa.

Also, strikingly, quantum oscillation experiments~\cite{terashima2010}
predicted effective charge carrier masses of up to $19 \, m_e$ with an
average mass enhancement factor $m^*/m_\text{band}$ of about 9, while
estimates from ARPES~\cite{yoshida2011} and cyclotron resonance
experiments~\cite{kimata2011} yield mass enhancements factors of about
3 for certain regions of the Fermi surface.  On the theoretical side,
density functional theory (DFT) calculations for {\kfeas} show poor
agreement to ARPES data~\cite{sato2009, yoshida2011, yoshida2012} and
to de Haas-van Alphen frequencies~\cite{terashima2010, terashima2013}.
However, an existing DFT+DMFT study~\cite{haulekotliar2011} for
{\kfeas} suggested an improved comparison of the Fermi surface contour
at $k_z$=0 with ARPES observations.

The importance of correlations in Fe-based superconductors has been
pointed out in the past by numerous studies. A method that has proven
to be quite successful in capturing the essential features of this
class of correlated metals is
DFT+DMFT~\cite{haulekotliar2011,Aichhorn2010,Hansmann2010,Aichhorn2011,Ferber1,Ferber2,Werner2012}.
Since an accurate knowledge of the electronic structure is essential
for understanding both the normal and the superconducting state in
Fe-based superconductors, we perform here a comprehensive LDA+DMFT
investigation focused on features of the {\kfeas} compound that have
not been dealt with in past studies\cite{haulekotliar2011}. We
critically benchmark our theoretical results with ARPES and dHvA
measurements to see in which way these results can improve our current
understanding of this system.

\section{METHODS}
\label{sec:methods}

We combine density functional theory with dynamical mean-field theory
in order to investigate the electronic structure and the resulting
dHvA frequencies of {\kfeas}.  Our calculations are based on the
experimentally determined tetragonal $I\,4/mmm$ structures of {\kfeas}
by Tafti {\it et al.}~\cite{tafti2014}, which are given for pressure
values starting at 0.23~GPa. We obtained a structure at zero pressure
by linear extrapolation of the available data points. A comparison of
this structure to the existing crystal structure by Rosza and
Schuster~\cite{roszaschuster1981} used in previous theoretical
investigations, shows that while lattice parameters $a$, $b$, and $c$
are consistent, the As~$z$ position differs significantly between both
structures. The As~$z$ position was consistently determined over a
large pressure range in the above mentioned study by Tafti {\it et
  al.}~\cite{tafti2014} and the As~$z$ position determined by Rosza
and Schuster doesn't follow the trend shown by those data.  Therefore
we use the new structure with the following parameters:
$a=b=3.8488$~{\AA}, $c=13.883$~{\AA}, fractional As~$z=0.140663$.

For the DFT calculation we employed the full-potential linear
augmented plane-wave (FLAPW) framework as implemented in
WIEN2k~\cite{wien2k}, using the local density approximation (LDA) as
well as the generalized gradient approximation (GGA) by Perdew, Burke
and Ernzerhof~\cite{perdewburkeernzerhof1996} to the
exchange-correlation functional.  These calculations were converged
self-consistently on a grid of 726 $k$ points in the irreducible
Brillouin zone. We performed calculations both, without and with
inclusion of spin-orbit coupling (SO).

In order to include the effect of local correlations we employed fully
charge self-consistent DMFT calculations, using modified routines from
the WIEN2K code. We considered our implementation of the projection method
from Bloch eigenstates to the correlated Fe $3d$ orbitals as described
by Aichhorn {\it et al.}~\cite{Aichhorn2009}. The energy window
for the projection onto the localized basis was chosen comparatively
large, ranging from $-5$~eV to $13$~eV to capture the higher energy
contribution of the Fe $3d$ orbitals to the density of states arising
from the hybridization with the As $4p$ orbitals.

The DMFT impurity problem was solved using the continuous-time
hybridization expansion quantum Monte Carlo solver in imaginary
frequency space as implemented in the ALPS~\cite{alps} code.  We
considered the Legendre polynomial representation~\cite{legendre} of
the Green's function and improved estimators for the
self-energy~\cite{improvedestimators}. About $6\times 10^6$
Monte-Carlo sweeps were performed at an inverse temperature
$\beta=40$~eV$^{-1}$, corresponding to room temperature. The
interaction parameters $U$ and $J$ were chosen as $U=4$~eV and
$J=0.8$~eV in terms of Slater integrals and for the double-counting
correction we used the fully-localized limit (FLL).  All orbital
characters presented here are defined in a coordinate system which is
rotated by $45^{\circ}$ around the crystallographic $z$ axis, {\it
  i.e.} $x$ and $y$ are pointing along Fe-Fe nearest neighbor bonds.
For determining the LDA+DMFT excitation energies that we used to
define the Fermi surface, we tracked the maximum of the real-frequency
spectral function throughout the Brillouin-zone.  Analytic
continuation of imaginary frequency data to the real frequency axis
was performed by using the Pad\'e approximation for the impurity
self-energy and we checked the results against the stochastic maximum
entropy method~\cite{beach2004}.

Two-dimensional cuts through the Fermi surface were extracted from
WIEN2k on $200 \times 200$ $k$ point grids for the LDA
calculations. The Fermi surface was interpolated linearly.

De Haas-van Alphen frequencies were calculated from the electronic
band structure using our own implementation of the dHvA frequency
extraction algorithm by Rourke and Julian~\cite{rourkejulian2012}.
Band energies were extracted on grids spanning the reciprocal unit
cell with 27648 $k$ points for the DFT(+SO) setups and 50000 $k$
points for the DMFT calculations.  The frequency calculation was
carried out on a $6.4 \cdot 10^7$ $k$ point supercell grid containing
64 reciprocal unit cells.  The higher $k$ point density in the
supercell compared to the input files is achieved by tricubic
interpolation~\cite{lekienmarsden2005}.  We align the supercell with a
fictitious magnetic field vector, which allows us to investigate the
angular dependence of dHvA frequencies. Here we varied the angle of
the magnetic field from the (001) towards the (010) direction in
reciprocal space.  In order to calculate frequencies, the
supercell is cut into one $k$ point thick slices perpendicular to the
vector of the magnetic field. On these slices a stepping algorithm
detects crossings of the Fermi level between grid points and
interpolates the position of the Fermi surface linearly. Point
ordering is ensured within the stepping algorithm. If a closed orbit
is detected, its area can be calculated from the geometry of the
points unambiguously. The three-dimensional Fermi surface is
reconstructed by matching orbits slice-by-slice. Only extremal orbits are
singled out since non-extremal orbits do not contribute in
experiment.  De Haas-van Alphen frequencies are calculated for the
extremal orbits and their positions are matched back to the reciprocal
unit cell. Orbits with similar position and frequency are averaged.

\section{RESULTS}
\label{sec:results}

\subsection{Electronic structure}
\label{subsec:electronic_structure}

\begin{figure}[tb]
\centering
\includegraphics[width=\linewidth]{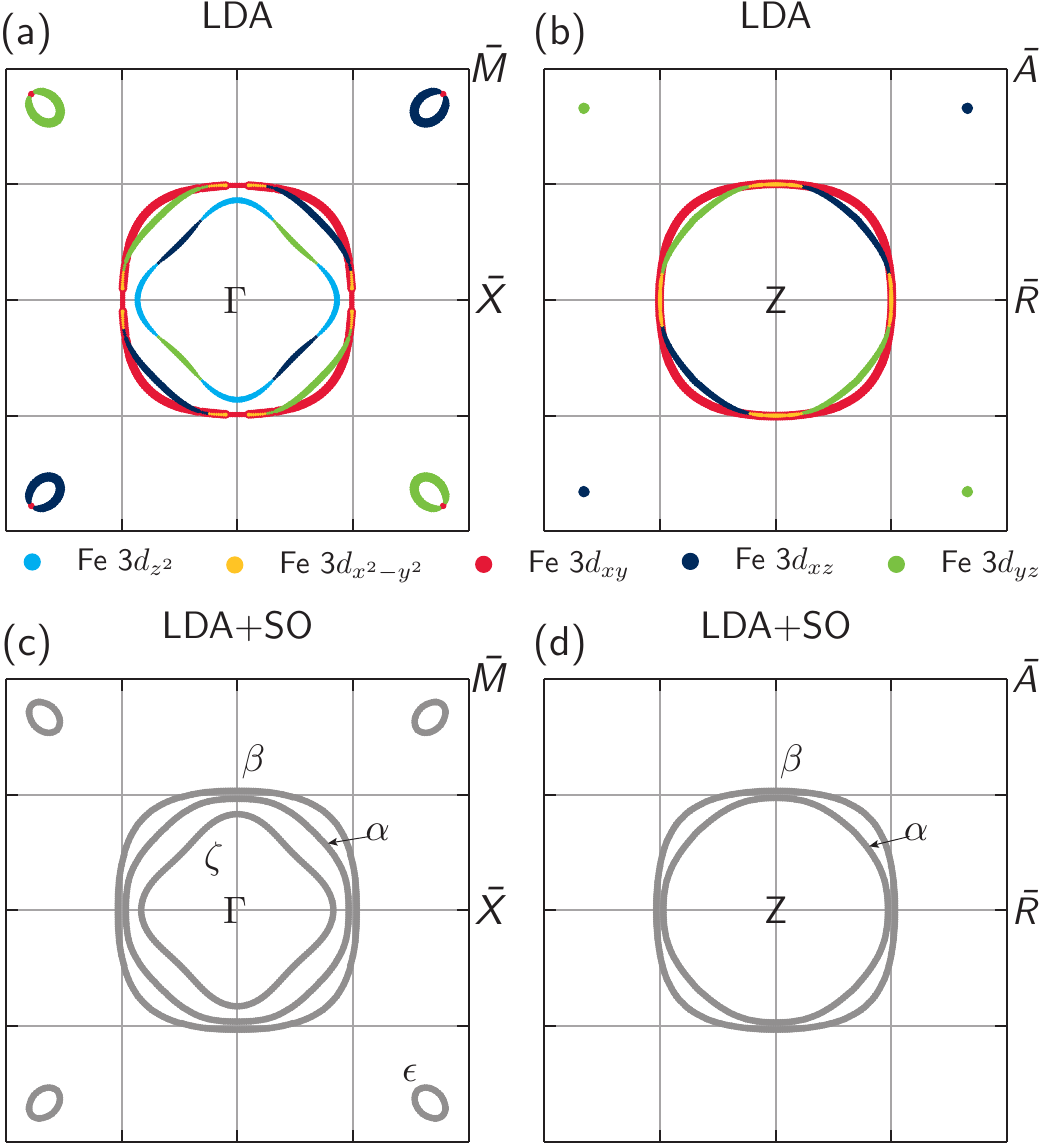}

\caption{(Color online) Overview of Fermi surface cuts at $k_z=0$ and
  $k_z=\pi$ in {\kfeas} obtained from DFT using the LDA exchange
  correlation functional with and without spin-orbit coupling
  (SO). Inclusion of spin-orbit coupling only leads to small
  quantitative changes, in particular a lifting of all apparent
  degeneracies of Fermi surface sheets. Fermi surfaces are shown in
  the two-Fe Brillouin zone representation.}
\label{fig:fsoverview}
\end{figure}

\begin{figure}[tb]
\begin{flushleft}
\includegraphics[width=0.46\textwidth]{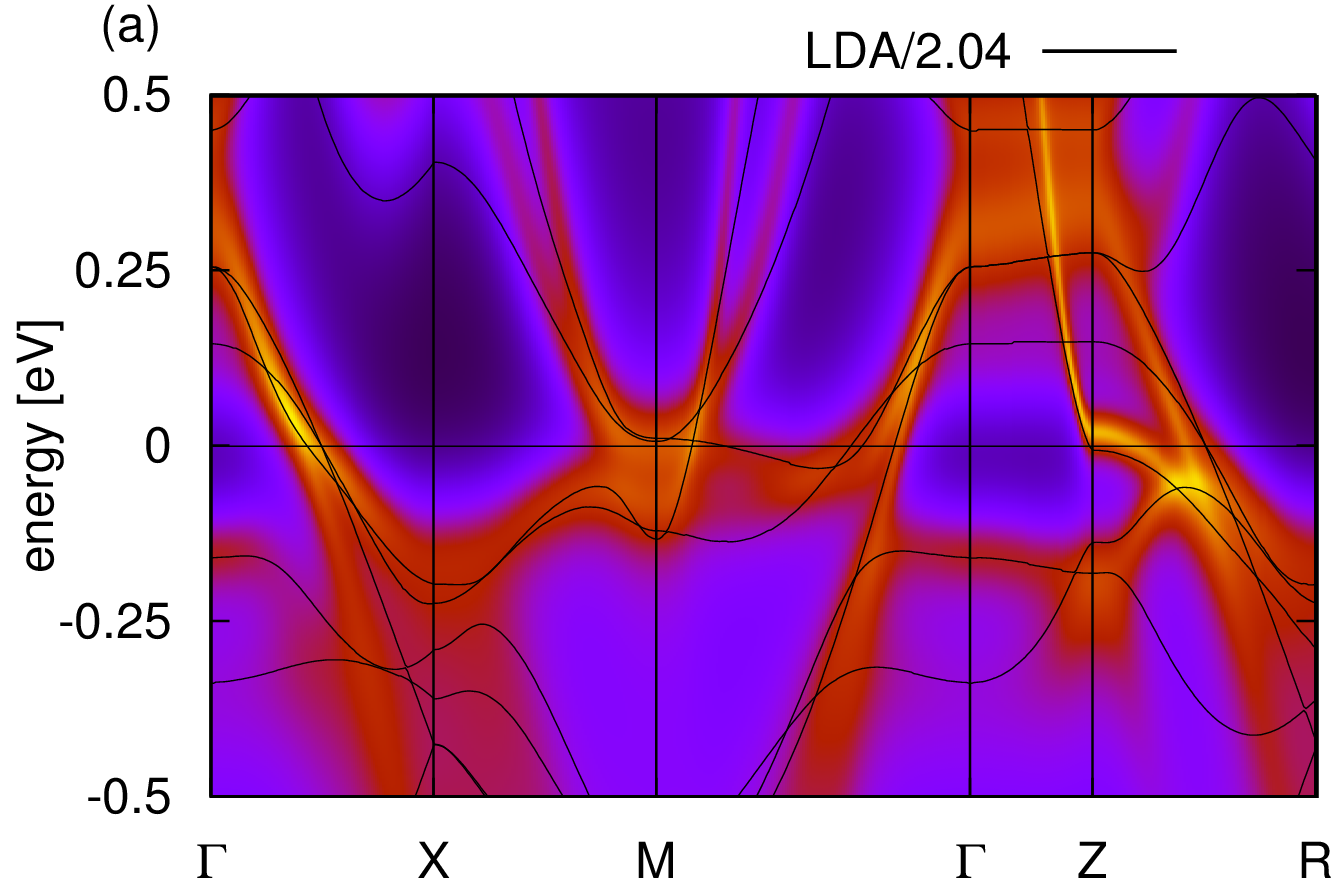} 
\end{flushleft}
\bigskip
\includegraphics[width=0.49\textwidth]{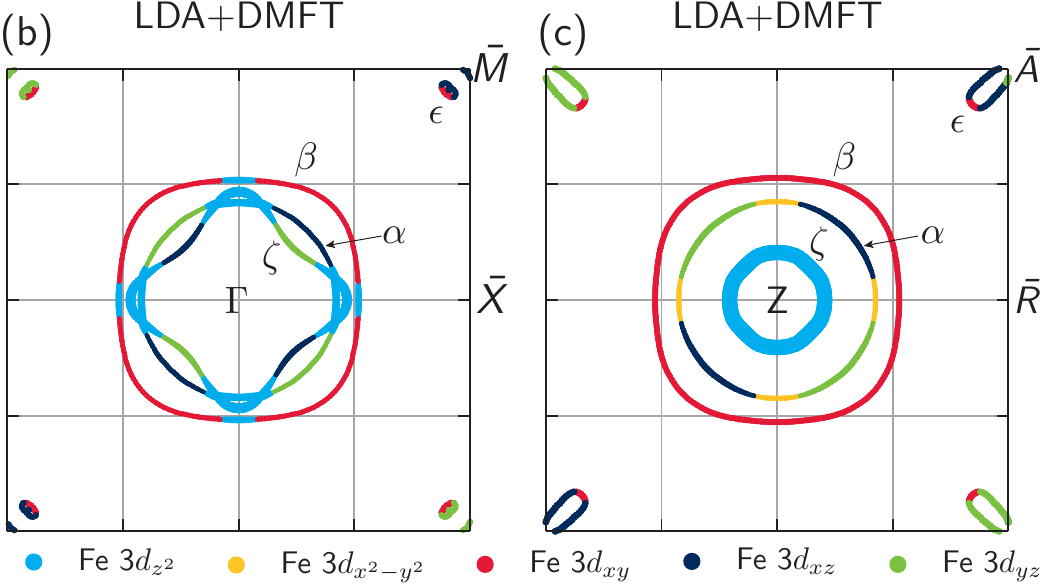}
\caption{(Color online) The $k$-resolved spectral function and
  orbital-resolved Fermi surface of {\kfeas} within LDA+DMFT (see
  detailed explanation in the text). The LDA bands (black lines) are
  rescaled by the average mass enhancement of 2.04 for comparison.  Dominant
  orbital characters are indicated by the color scale. Fermi surfaces
are shown in the two-Fe Brillouin zone representation.}
\label{fig:gga_dmft_fs}
\end{figure}

We first investigated the band structure of {\kfeas} obtained by the
DFT calculation within LDA and LDA+SO. Results obtained with the GGA
functional were nearly identical to the LDA result and are therefore
not shown. At the $\Gamma$ point (see Fig.~\ref{fig:fsoverview} (a)
and Fig.~\ref{fig:gga_dmft_fs} (a)) we see three bands crossing the
Fermi level, forming hole pockets of Fe $3d_{xy}$, $3d_{xz}$ and
$3d_{yz}$ character.  The two outer hole pockets form cylinders
between the $\Gamma$ and $Z$ points, while the cylinder of the third
inner hole pocket closes shortly before $Z$. This leads to two hole
pockets at the $Z$ point, being mostly of Fe $3d_{xy}$, $3d_{xz/yz}$
character.  Around the $\bar{M}$ point, we observe very small
hole pockets with Fe $3d_{xy}$, $3d_{xz/yz}$ character, where the
bands with mostly Fe $3d_{xz/yz}$ character are very shallow right
above $E_F$, which leads to a high sensibility to input parameters and
total electron charge in the calculation.

By including the spin-orbit interaction we observe an overall
repulsion between touching or degenerate bands, which leads to clear
separation of the hole pockets along the high symmetry directions (see
Fig.~\ref{fig:fsoverview} (c), (d)).

Comparing these DFT results to ARPES measurements~\cite{sato2009,
  yoshida2011, yoshida2012} we find that the agreement in size and
shape of the hole pockets along the high symmetry directions is quite
poor.  This disagreement has already been noted in the above cited
publications. The inner two pockets ($\alpha$, $\zeta$) are too large
while the outer one ($\beta$) is too small. The topology of the Fermi
surface also differs from experimental observations.  ARPES clearly
shows a separated outer hole cylinder at $\Gamma$, while the two inner
ones overlap considerably~\cite{sato2009, yoshida2011, yoshida2012}.
The closure of the inner hole cylinder is not seen in
ARPES~\cite{sato2009, yoshida2011, yoshida2012} or
dHvA~\cite{terashima2010, terashima2013} measurements, leading to a
third inner hole pocket at $Z$ in experiments. Also, the hole pockets
($\epsilon$) close to $\bar{ \mathrm{M}}$ are too small in DFT.

When including correlations on the Fe $3d$ orbitals via DMFT, the
electronic structure of {\kfeas} changes significantly.  Dynamical
mean-field theory yields very similar results for both LDA and GGA
initial calculations of the electronic structure. In what follows we
present the LDA+DMFT results.  In the band structure shown in
Fig.~\ref{fig:gga_dmft_fs} we observe a strong renormalization of the
bands around the Fermi level, with mass enhancements for the Fe $3d$
orbitals ranging from $1.56$ to $2.72$, as shown in
Table~\ref{tab:effmasses}.

These results are in agreement with a previous LDA+DMFT
study~\cite{haulekotliar2011}, but still very different from
experimentally reported mass enhancements, that can reach values of up
to 24 for the small pockets ($\epsilon$) and up to 6.9 for the large
pockets ($\alpha$, $\beta$, $\zeta$) in the center of the reciprocal
unit cell~\cite{terashima2010, terashima2013}.  It was pointed out in
Ref.~\cite{terashima2013} that flattening of the bands near the Fermi
level caused by coupling to low-energy bosonic
excitations~\cite{ortenzi2009} also contributes to mass enhancements,
but may not be accounted for in the DMFT method.

\begin{table}[b]
\begin{ruledtabular}
\begin{tabular}{ccccc}
Orbital &  $d_{xy}$ & $d_{z^2}$ & $d_{x^2-y^2}$ & $d_{xz/yz}$  \\
\hline
$\frac{m^*}{m_{LDA}}$ & 2.72 & 1.89 & 1.56 & 2.02 \\
\end{tabular}
\end{ruledtabular}
\caption{The orbital-resolved mass enhancements for the Fe $3d$ orbitals in {\kfeas}.}
\label{tab:effmasses}
\end{table}

\begin{figure}[tb]
\subfigure[]{\includegraphics[height=0.2\textwidth]{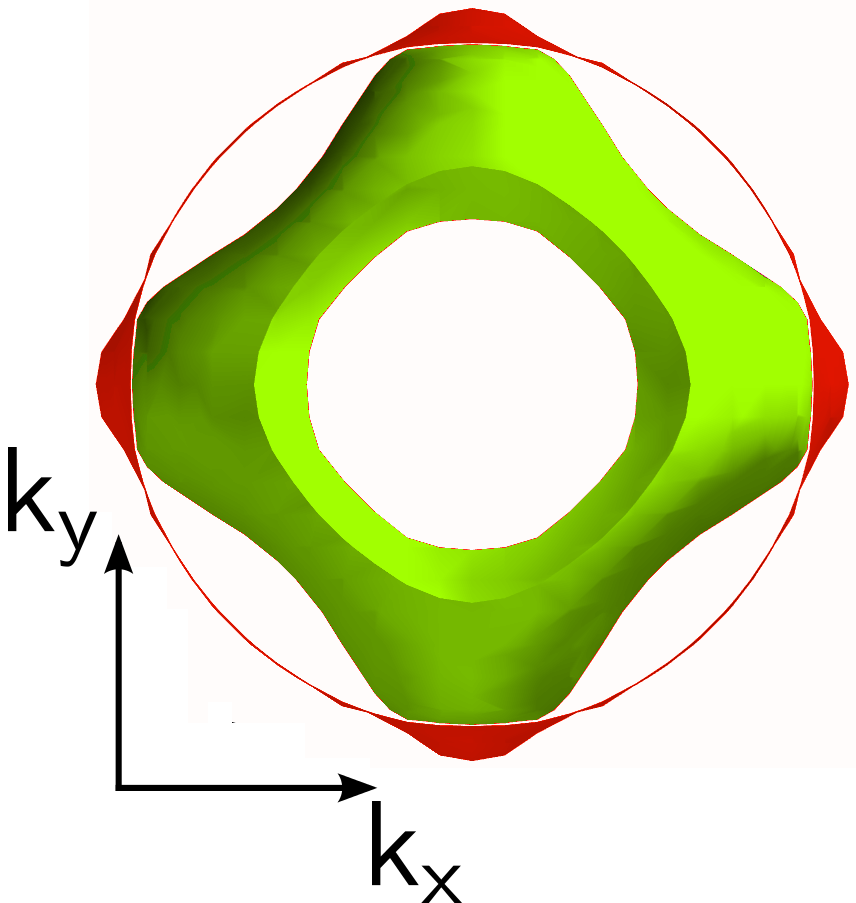}}
\hspace{3em}
\subfigure[]{\includegraphics[height=0.2\textwidth]{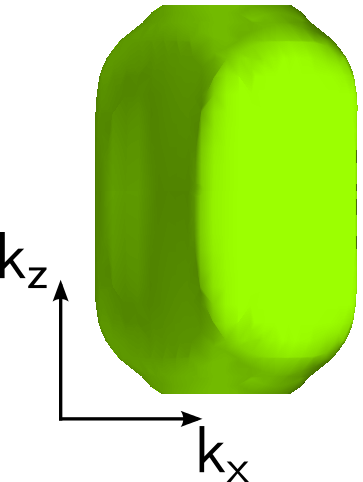}}
\caption{(Color online) Three dimensional view of the Fermi surface
  obtained from LDA+DMFT in the two-Fe Brillouin zone
  representation. Figure (a) shows the intersection nodes between the
  inner (green) and middle (red) Fermi surface sheet. The
  configuration used in calculating de Haas-van Alphen frequencies is
  indicated by the colors. Figure (b) shows the dispersion of the
  inner Fermi surface sheet along the $k_z$-axis.}
\label{fig:fs3dnodes}
\end{figure}

\begin{figure}
\includegraphics[width=\linewidth]{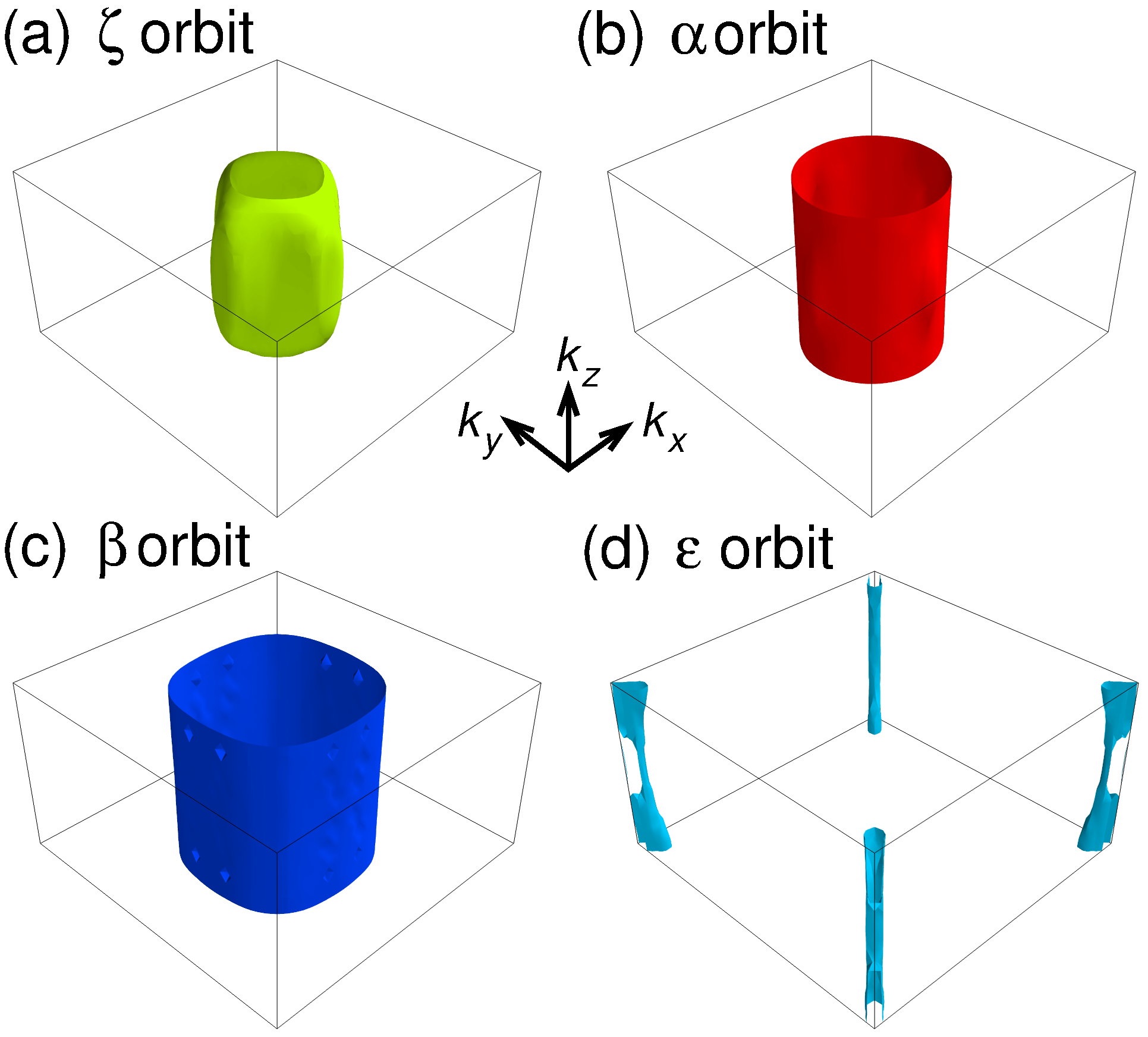}
\caption{Three dimensional view of the Fermi surface obtained from
  LDA+DMFT in the two-Fe Brillouin zone representation. Figure (a)
  shows the sheet that we attribute to the $\alpha$ orbit observed in
  de Haas-van Alphen experiments. Figure (b) shows the cylinder
  attributed to the $\zeta$ orbit. Figure (c) shows the cylinder
  attributed to the $\beta$ orbit. The small cusps are artifacts from
  the technical procedure that we carefully exclude in de Haas-van
  Alphen calculations. Figure (d) shows the sheets that we attribute
  to the $\epsilon$ orbits.}
\label{fig:fs3d}
\end{figure}

We also observe a reordering of bands along the high symmetry
directions with significant changes in the size of the hole cylinders.
Both the inner sheets ($\alpha$, $\zeta$) at the $\Gamma$ point shrink
in size, while the outer one ($\beta$) gets enlarged, as seen in the
LDA+DMFT Fermi surface in Fig.~\ref{fig:gga_dmft_fs}.  This is in
better agreement to the experimental observations.  Moreover, we
observe a small overlap of the center ($\zeta$) and middle hole pocket
($\alpha$) with small intersection nodes around $\Gamma$, which are
also observed in ARPES but were absent in the DFT calculation and
previous LDA+DMFT studies~\cite{haulekotliar2011}.  Most importantly,
at the $Z$-point the band of mostly 3$d_{z^2}$ character that was
located just below the Fermi level is pushed above $E_F$ due to
correlations, opening the hole cylinder that was previously closed in
the DFT calculation. By also investigating the structure from Rosza
and Schuster~\cite{roszaschuster1981}, we found a strong dependence of
the shape of this hole pocket on the As-$z$ position (see Appendix).
 Within DFT
alone, an opening of a new hole pocket can be observed by increasing
the As height above the Fe plane. Since the band in question
originates from the hybridization of Fe $3d$ with As $4p$ states, it
is extremely sensitive to the arsenic position.  The LDA+DMFT middle
hole pocket around $Z$ reduces in size compared to LDA, forming an
almost $k_z$-dispersionless hole cylinder between the $\Gamma$ and $Z$
points.  In Figs.~\ref{fig:fs3dnodes} and \ref{fig:fs3d} we show
three-dimensional plots of the hole cylinders throughout the Brillouin
zone.

Recent ARPES experiments~\cite{sato2009, yoshida2011, yoshida2012} and
dHvA measurements~\cite{terashima2010, terashima2013} also observe
three hole pockets at the $Z$ point, agreeing well with our
calculations.  The strong Fe $3d_{z^2}$ character around Z reported
from ARPES~\cite{yoshida2012} is also reproduced by our calculation.
A detailed comparison shows, however, still
 some differences between theory and ARPES experiment:
 the size of the middle hole pocket at both the $\Gamma$ and
$Z$ points is smaller in ARPES, while the inner pocket at $Z$ seems to
be larger compared to our results.

The small hole pockets at the $\bar{M}$ point emerge from the crossing
of two bands at an energy of about 5~meV above $E_F$, with very weak
dispersion of one of the bands.  Therefore, these pockets are
extremely sensitive to the Fermi level which makes them strongly
dependent on the details of the calculation like the double-counting
scheme or the chosen DFT functional. On the experimental side this
indicates a strong dependence on the actual composition and possible
impurities in the sample; this is a possible explanation for the
different sizes of these pockets in ARPES experiments~\cite{sato2009,
  yoshida2011, yoshida2012}.  
In our calculations, we carefully
checked the results for different double-counting procedures and
analytic continuation methods and found overall qualitatively good
agreement.  Finally, we note that in order to obtain a better
agreement with experiment, the middle hole cylinder would have to be
shifted inside of the inner cylinder in our calculations.  This cannot
be achieved by inclusion of local correlations only.

The Fermi surface obtained from LDA+DMFT offers a natural
explanation for the magnetic breakdown junctions between orbits
$\alpha$ and $\zeta$ observed by Terashima et al.~\cite{terashima2010,
  terashima2013}.  Therefore we conclude that the degeneracy of the
lines found in ARPES might be due to both experimental resolution and
overestimation of the distance between sheets in our calculation.
Furthermore we would like to point out that our Fermi surface strongly
resembles the octet line-node structure observed in laser ARPES
measurements of the superconducting order
parameter~\cite{okazaki2012}.

\subsection{De Haas-van Alphen effect}
\label{subsec:dhva_effect}

\begin{figure*}[t]
\centering
\begin{tabular}{ccc}

\subfigure{\includegraphics[width=0.24\textwidth]{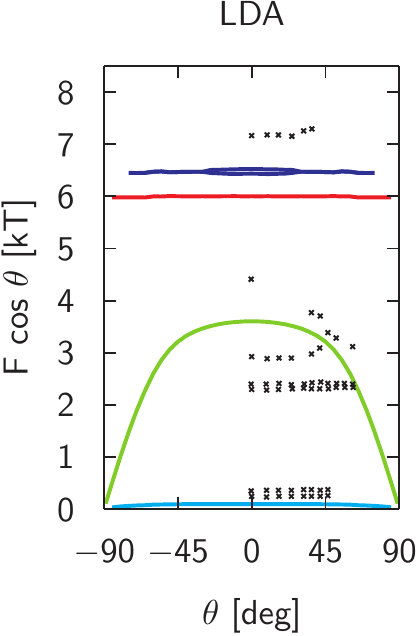}} &
\subfigure{\includegraphics[width=0.24\textwidth]{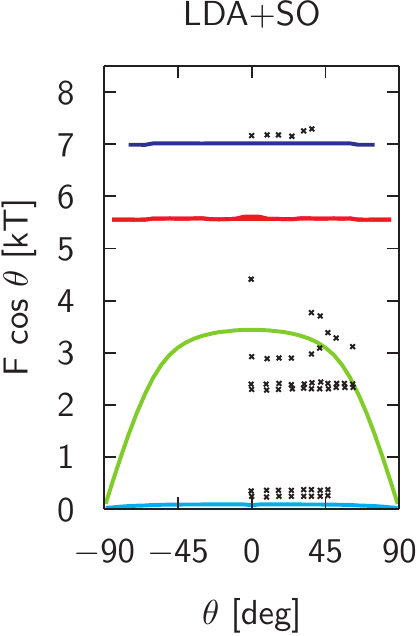}} &
\subfigure{\includegraphics[width=0.24\textwidth]{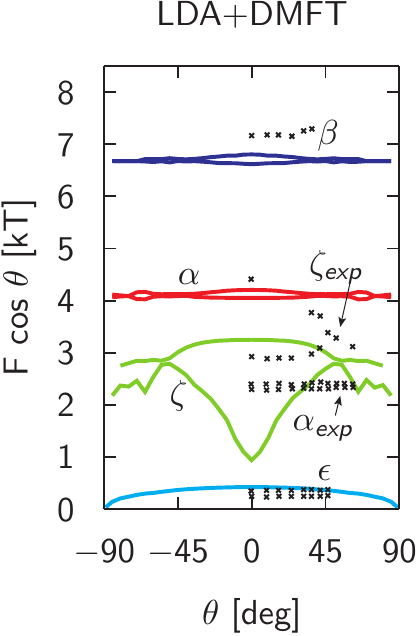}} \\
\end{tabular}
\caption{(Color online) Overview of de Haas-van Alphen frequencies in
  {\kfeas} calculated from density functional theory and dynamical
  mean-field theory. Lines represent our calculations, while crosses
  represent experimental frequencies taken from
  Ref.~\onlinecite{terashima2013}. Color coding is the same as in
  Fig.~\ref{fig:fs3d}. The $\zeta$-orbit (innermost) is shown in
  green, while the frequencies originating from the middle sheet
  ($\alpha$) are shown in red. The outermost orbits ($\beta$,
  $\epsilon$) are drawn in blue.}
\label{fig:dhvaoverview}
\end{figure*}

Comparing our findings to measurements of quantum
oscillations~\cite{terashima2013} we can confirm that DFT is not able
to describe the Fermi surface of {\kfeas} correctly. An overview of
our results is presented in Fig.~\ref{fig:dhvaoverview}. LDA and
LDA+SO calculations for the structure by Rosza and Schuster are given
in the Appendix. They reproduce the DFT results by Terashima {\it et
  al.}~\cite{terashima2010, terashima2013}.

The two inner hole pockets ($\alpha$, $\zeta$) around the $\Gamma$
point are too large compared to experimental frequencies, while the
outermost hole pocket ($\beta$) is too small 
(Fig.~\ref{fig:dhvaoverview} left panel). The size of the hole
pocket close to $\bar{ \mathrm{M}}$ ($\epsilon$) is already well
described in DFT.  Adding spin-orbit coupling already shows the
correct tendency to increase the size of the outer hole pocket and
decrease the size of the two inner hole pockets. Deviations from
experimentally observed frequencies are nevertheless large
(Fig.~\ref{fig:dhvaoverview} middle panel). The good
agreement with experiment for the largest and smallest frequencies
comes with persisting disagreement for the two intermediate
frequencies.

In the LDA+DMFT calculation (Fig.~\ref{fig:dhvaoverview} right panel)
the two innermost orbits
($\alpha$,$\zeta$) intersect around the $\Gamma$ point
(Fig.~\ref{fig:gga_dmft_fs}).  For the analysis of the dHvA
frequencies we take into account the outermost and innermost possible
configuration of these two orbits as shown in Fig.~\ref{fig:fs3d}. The
same configuration was attributed to fundamental frequencies observed
in dHvA experiment~\cite{terashima2010, terashima2013}.  The outer
hole pocket ($\beta$) is considerably enlarged. As the corresponding
electronic band is flattened, it becomes susceptible to tiny energy
shifts. Both inner hole pockets ($\alpha$, $\zeta$) are shifted to
lower frequencies and thus decreased in size. The small orbit
($\epsilon$) close to $\bar{ \mathrm{M}}$ is enlarged around the $Z$
point, but decreases in size around $\Gamma$ as shown in
Fig. \ref{fig:gga_dmft_fs}. Therefore we only find the maximum
frequency for this sheet.

We would like to note that quantum oscillation
experiments~\cite{zocco2014} and ARPES\cite{yoshida2011, yoshida2012} reported the existence of a fourth very
small pocket centered at the $Z$ point which was not seen in our
calculations for the most recent structure. This fourth pocket is
however present in the DFT calculation when using the structure from
Rosza and Schuster~\cite{roszaschuster1981}, but we found it to vanish
when adding correlations in LDA+DMFT, depending on the double-counting (see Appendix).

The closing of the innermost hole pocket $\zeta$ is clearly observed
in the LDA+DMFT calculated dHvA frequencies by the appearance of a
lower extremal frequency. As pointed out before in the ARPES section,
the middle hole cylinder would have to be decreased in size
considerably to match the experimental frequencies. This would in turn
increase the enclosed volume of the sheet labeled $\zeta$ and thus
shift it toward experimentally observed values.
A comparison of experimental and LDA+DMFT frequencies for $B \parallel (001)$
is given in Table~\ref{tab:dhvafrequencies}.

\begin{table}[tb]
\begin{ruledtabular}
\caption{De Haas-van Alphen frequencies in kT (kiloTesla) for
$B \parallel (001)$ obtained from DMFT calculations compared to
experimental values~\cite{terashima2013}.}
\label{tab:dhvafrequencies}
\begin{tabular}{rrrrrrrrr}
& $\epsilon_l$ & $\epsilon_h$ & $\alpha_l$ & $\alpha_h$ & $\zeta_l$ & 
$\zeta_h$ & $\beta_l$ & $\beta_h$ \\
\hline
exp. & 0.24 & 0.36 & 2.30 & 2.39 & 2.89 & 4.40 & 7.16 & - \\
LDA+DMFT & - & 0.42 & 4.05 & 4.20 & 0.94 & 3.25 & 6.62 & 6.81 \\
\end{tabular}
\end{ruledtabular}

\end{table}
\begin{table}[tb]
\begin{ruledtabular}
\caption{Electron orbit averaged effective masses in $m_e$ for
$B\parallel (001)$ obtained from DMFT calculations compared to
experimental values~\cite{terashima2013}.}
\label{tab:dhvamasses}
\begin{tabular}{rrrrrrrrr}
& $\epsilon_\Gamma$ & $\epsilon_Z$ & $\alpha_\Gamma$ & $\alpha_Z$ & 
$\zeta_\Gamma$ & $\zeta_Z$ & $\beta_\Gamma$ & $\beta_Z$ \\
\hline
exp. & 6.0 & 7.2 & 6.0 & 6.5 & 8.5 & 18.0 & 19.0 & 19.0 \\
LDA+DMFT & - & 5.9 & 3.4 & 4.6 & 2.4 & 5.3 & 8.3 & 8.3 \\
\end{tabular}
\end{ruledtabular}
\end{table}

Furthermore we calculated effective masses averaged over extremal
orbits on the Fermi surface from the LDA+DMFT excitation
energies. These masses correspond to the effective masses observed in
dHvA experiments (Table~\ref{tab:dhvamasses}). Note that values given
in this table are \textit{absolute} masses in contrast to mass
\textit{enhancements} given in Table~\ref{tab:effmasses}.
 
Qualitatively our calculation captures the trends that are observed in
sheet-resolved effective masses, however we certainly miss effects
originating from other than electron-electron interactions, which
increase effective masses seen in dHvA experiments such as
electron-phonon coupling.

\section{CONCLUSION}
\label{sec:conclusion}
In this paper we presented combined density functional theory with
dynamical mean-field theory calculations of the Fermi surface and de
Haas-van Alphen frequencies in {\kfeas}. We first showed that DFT
calculations with LDA or GGA exchange correlation functionals, with or
without spin-orbit coupling fail to reproduce the experimentally
observed electronic structure of {\kfeas}.

Most notably, DFT predicts no third inner hole pocket at the $Z$
point, which we find to open in our LDA+DMFT calculation in agreement
with experiment.  We also obtain a qualitatively correct $k_z$
dispersion of the iron bands, where between the $\Gamma$ and $Z$
points the dispersion of the inner hole cylinder is greatly increased
with almost no dispersion of the middle hole cylinder, giving a much
better agreement with dHvA measurements when identifying them in
different order in experiment.

The intersection nodes we found on the inner two hole cylinders offer
a natural explanation for magnetic breakdown orbits observed in dHvA
measurements (Ref.~\onlinecite{terashima2013}).

The obtained effective mass-enhancements about $1.6-2.7$ show that
{\kfeas} is a moderately correlated metal and thus a DFT calculation
fails to capture the important features that lead to the
experimentally observed electronic structure. This has strong
implications for the obtained dHvA frequencies, where LDA+DMFT gives
distinctively different results than DFT.  Our results are in better
agreement with both ARPES (Refs.~\onlinecite{yoshida2011,yoshida2012})
and quantum oscillation (Refs.~\onlinecite{terashima2010,
  terashima2013}) experiments. The observed strong flattening of
electronic bands gives a possible explanation for the spread of
experimental results in this compound in terms of extreme sensitivity
to the experimental stoichiometry. 
 We conclude that LDA+DMFT
captures most of the important correlation effects in {\kfeas}
and such a treatment may be necessary in order to understand 
the controversial nature of superconductivity in this system.
 This will be a subject of future work.

\acknowledgments 

The authors would like to thank Milan Tomi\'c, Markus Aichhorn,
Emanuel Gull, Luca de Medici, Peter J. Hirschfeld,
Amalia Coldea and Kristjan Haule for useful
discussions and suggestions and gratefully acknowledge financial support by the
Deutsche Forschungsgemeinschaft through grant SPP 1458.  Daniel
Guterding acknowledges support by the German National Academic
Foundation.

\begin{appendix}

\section{Sensitivity analysis of the Fermi surface and de Haas-van Alphen frequencies}
Previous theoretical works on {\kfeas} were based on the structural
data obtained by Rosza and Schuster~\cite{roszaschuster1981}, while
new results also for higher pressures became recently available by
Tafti {\it et al.}~\cite{tafti2014}.  These structures differ most noticeably
in the As $z$-position, where the old structure has a fractional coordinate 
of $z=0.1475$, while the new one yields $z=0.140663$ by interpolation to
0~GPa.  Therefore, to interpret current theoretical investigations
correctly, we investigate the dependence of the Fermi surface and de
Haas-van Alphen frequencies of {\kfeas} on the two different
structures and also on different double-counting methods within
LDA+DMFT. We also tested the dependence of these quantities
upon considering  LDA versus GGA
and found very minor changes. Here we present GGA results.

\begin{figure}
\includegraphics[width=\linewidth]{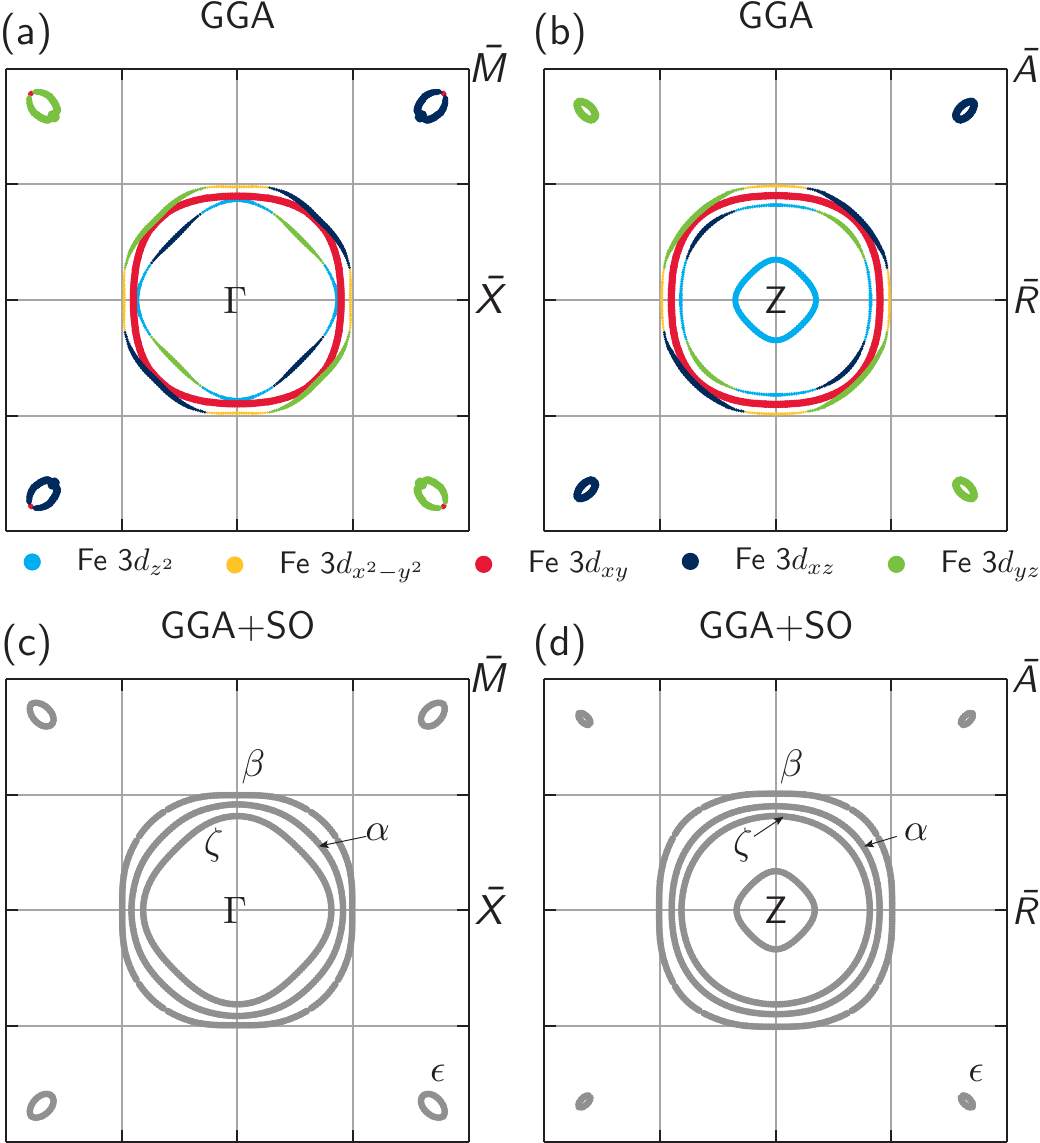}
\caption{(Color online) The orbital-resolved Fermi surface of {\kfeas}
  within DFT for the structural data by Rosza and
  Schuster~\cite{roszaschuster1981}. Dominant orbital characters are
  indicated by the color scale. Fermi surfaces are shown in the two-Fe
  Brillouin zone representation.}
\label{fig:fsdftold}
\end{figure}

We find very different behavior for the two structural
configurations: The Fermi surfaces for the Rosza and
Schuster~\cite{roszaschuster1981} structure can be seen in
Fig.~\ref{fig:fsdftold}, to be compared to the
 Tafti {\it et al.}~\cite{tafti2014} structure in Fig.~\ref{fig:fsoverview}. The cut
at $k_z=0$ is qualitatively identical but the Fermi surface topology
at the $Z$-point is different, where the structure of Rosza and
Schuster with the higher As $z$-position features two additional inner
hole pockets around $Z$. The inner one emerges from a small hole
pocket centered at $Z$, while the second inner one corresponds to the
open $\zeta$-hole cylinder, which closes shortly before $k_z=\pi$ in
the structure by Tafti {\it et al.}~\cite{tafti2014}.
  From our calculations we can deduce
that the As $z$-position is the key factor for the existence of these
hole pockets.  This makes sense since their main orbital character is
either Fe $3d_{z^2}$ or As $4p$, giving rise to a strong dependence on
the Fe-As bonding distance.  By lowering the As $z$-position, and thus
enhancing the hybridization of the Fe $3d$ with the As $4p$ orbitals,
the two inner pockets become smaller and finally vanish.

\begin{figure}
\subfigure{}{\includegraphics[width=\linewidth]{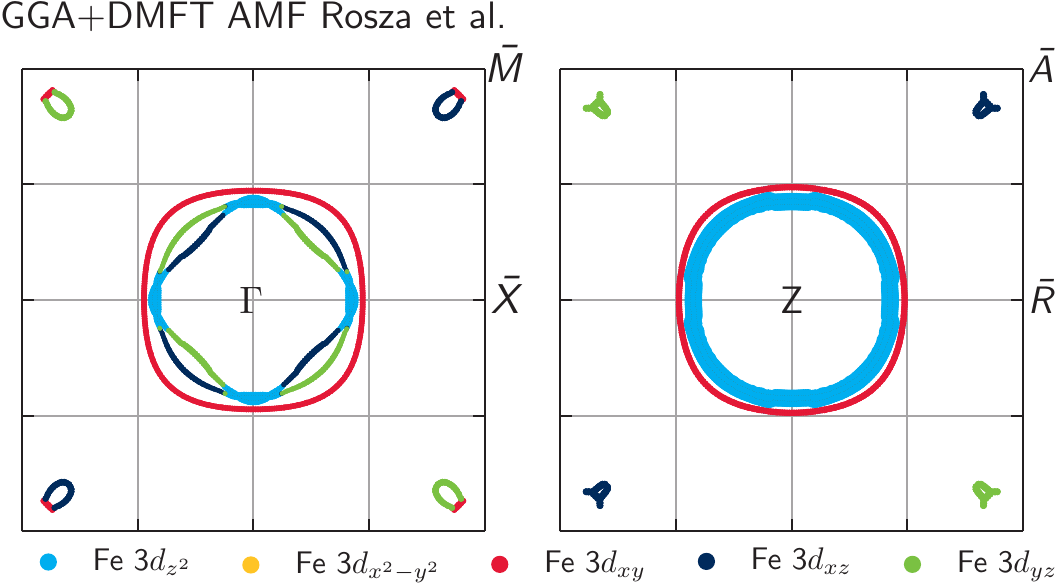}}
\subfigure{}{\includegraphics[width=\linewidth]{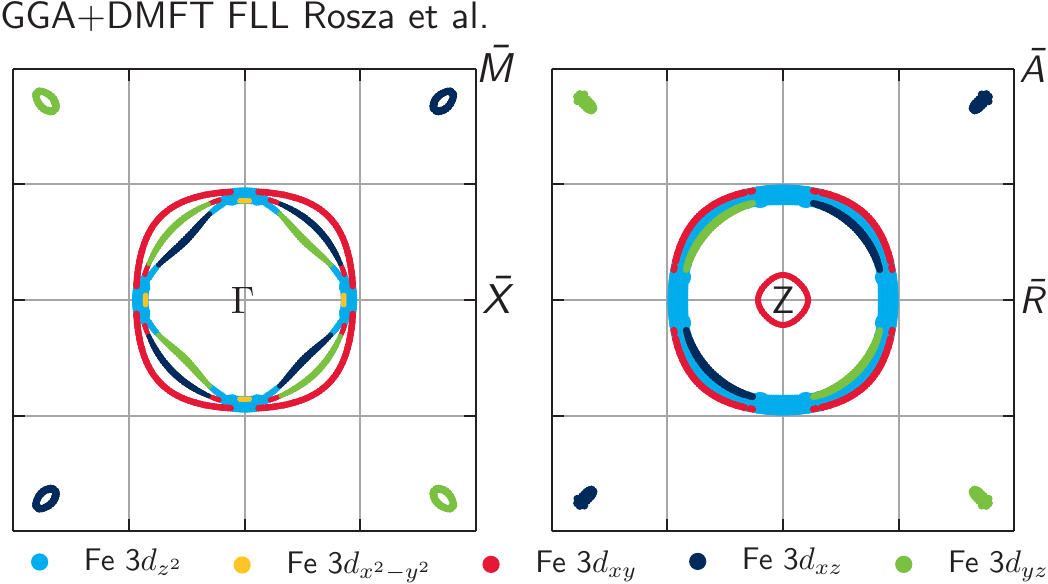}}
\caption{(Color online) The orbital-resolved Fermi surface of {\kfeas}
  within GGA+DMFT using the structural data by Rosza and
  Schuster~\cite{roszaschuster1981}. Dominant orbital characters are
  indicated by the color scale. Fermi surfaces are shown in the two-Fe
  Brillouin zone representation. AMF indicates the around mean-field,
  FLL the fully-localized limit double-counting.}
\label{fig:fsdmftold}
\end{figure}

In Fig.~\ref{fig:fsdmftold} we show cuts of the Fermi surface in
LDA+DMFT, calculated for the structure by Rosza and Schuster.  The
electronic structure depends on the double-counting correction, where
with the FLL double-counting the fourth inner hole pocket stays present
at the $Z$ point, whereas with the around mean-field (AMF) double-counting it
vanishes. This can be explained by the large As $4p$ character of the
inner hole pocket, which makes it sensible to the double-counting
method. Since AMF reduces the self-energy by a smaller degree than
FLL, this band is pushed below the Fermi level when using the AMF
method.  In the structure by Tafti {\it et al.} this band is  further
below the Fermi level already in the DFT calculation, lowering the As
$4p$ contribution to the density of states at $E_F$. Therefore, we see
only slight differences between the two double-counting corrections
in the Tafti {\it et al.} structure with no
qualitative changes.

\begin{figure*}
\centering
\begin{tabular}{ccc}

\subfigure{\includegraphics[width=0.24\textwidth]{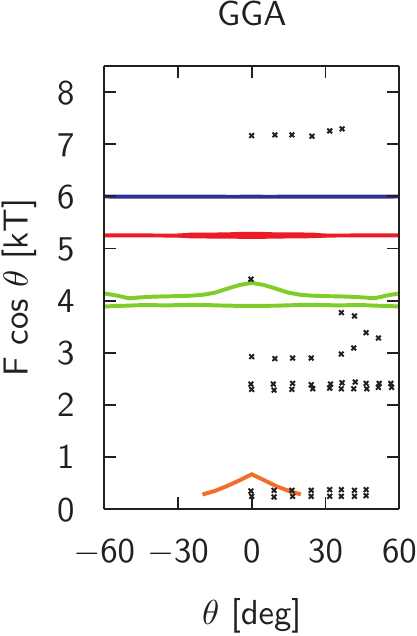}} &
\subfigure{\includegraphics[width=0.24\textwidth]{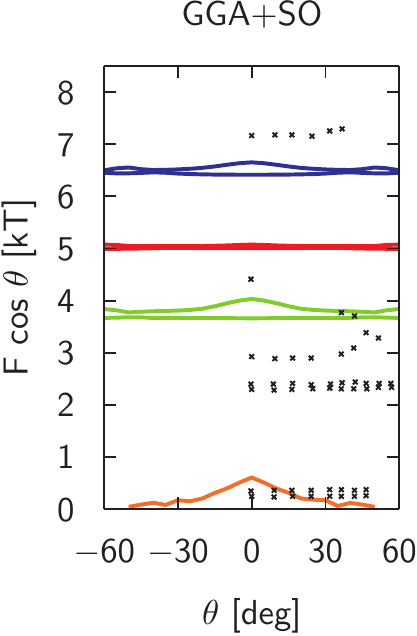}} &
\subfigure{\includegraphics[width=0.24\textwidth]{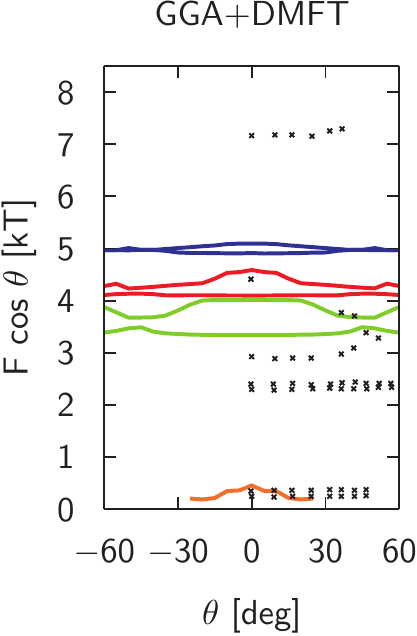}} \\
\end{tabular}
\caption{(Color online) Overview of de Haas-van Alphen frequencies in
  {\kfeas} calculated from density functional theory and dynamical
  mean-field theory with FLL double-counting using the structural data by Rosza and
  Schuster~\cite{roszaschuster1981}. 
  Lines represent our calculations, while crosses represent
  experimental frequencies taken from
  Ref.~\onlinecite{terashima2013}. Color coding is the same as in
  Fig.~\ref{fig:fs3d}. The $\zeta$-orbit (innermost) is shown in
  green, while the frequencies originating from the middle sheet
  ($\alpha$) are shown in red. The outermost orbits ($\beta$,
  $\epsilon$) are drawn in blue. The orbit shown in orange corresponds
  to a small pocket at the $Z$-point.}
\label{fig:dhvadmftold}
\end{figure*}

In Fig.~\ref{fig:dhvadmftold} we show the dHvA frequencies 
for the structure by Rosza and Schuster. The small orbits
at the $\bar M$-\- point are not shown as they 
give very low frequencies.

\end{appendix}

\end{document}